\documentclass{Interspeech2024}

\interspeechcameraready

\title{A Joint Noise Disentanglement and Adversarial Training Framework for Robust Speaker Verification}

\name[affiliation={1}]{Xujiang}{Xing}
\name[affiliation={2}]{Mingxing}{Xu}
\name[affiliation={2*}]{Thomas Fang}{Zheng}

\address{
  $^1$School of computer science and technology, Xinjiang University, China\\
  $^2$Tsinghua University, China\thanks{*Corresponding author}
  }

\email{xingxj@stu.xju.edu.cn, \{xumx,fzheng\}@tsinghua.edu.cn}

\keywords{speaker verification, noise-robust, multi-task, adversarial training}

\usepackage{multirow}
\begin{document}

\maketitle

\begin{abstract}
    Automatic Speaker Verification (ASV) suffers from performance degradation in noisy conditions. To address this issue, we propose a novel adversarial learning framework that incorporates noise-disentanglement to establish a noise-independent speaker invariant embedding space. Specifically, the disentanglement module includes two encoders for separating speaker related and irrelevant information, respectively. The reconstruction module serves as a regularization term to constrain the noise. A feature-robust loss is also used to supervise the speaker encoder to learn noise-independent speaker embeddings without losing speaker information. In addition, adversarial training is introduced to discourage the speaker encoder from encoding acoustic condition information for achieving a speaker-invariant embedding space. Experiments on VoxCeleb1 indicate that the proposed method improves the performance of the speaker verification system under both clean and noisy conditions.
\end{abstract}
\vspace{-0.2cm}
\section{Introduction}

Automatic speaker verification (ASV) aims to verify the identity of the speaker using their voice \cite{10053562}. The most advanced speaker recognition systems \cite{zeinali2019but, desplanques20_interspeech, zhang22h_interspeech} can achieve remarkable performance under acoustic control conditions. However, in real environments, the degradation of speech signals caused by background noise can significantly reduce the performance of speaker recognition systems \cite{9379245}. That is due to that noise can disrupt the voiceprint characteristics of clean speech and cause a distribution mismatch between test and training speech, which is typically devoid of noise.

In the last years, extensive research was conducted on reducing the adverse effects of noise on speaker recognition systems \cite{peri2020robust, li21b_interspeech, wu21c_interspeech}. One method is to extract noise-robust speaker embedding by reducing the embedding distance between noisy/clean pairs. MohammadAmini et al. \cite{mohammadamini2022learning} proposed an optimal training strategy to make the extracted x-vector in noisy environments close to the corresponding x-vector in clean environments. Traditional speech enhancement (SE), aiming to improve speech quality by suppressing noise, may be detrimental to speaker verification \cite{shon19b_interspeech}. Unlike traditional SE, joint training of speaker recognition systems and front-end enhancement modules is a novel approach \cite{kim22b_interspeech, 10122599}. Han et al. \cite{10201239} utilized the combined model of SE and speaker verification as a pre-trained model to extract noise-robust embedding. And some other methods for extracting robust speaker embeddings. Yu et al. \cite{9414704} presented context-aware masking to extract robust speaker embedding by enabling the neural network to focus on the speaker of interest and blur irrelevant noise. 

Data augmentation is also one of the most commonly used methods to improve the robustness of speaker recognition systems. Wang et al. \cite{10095066} proposed a novel difficulty-aware semantic augmentation approach for generating diverse training samples at the speaker embedding level. Joint training of speaker recognition systems using clean and noisy data often yields satisfactory results \cite{8461375}, but the performance of the SV system degrades sharply when facing unseen noises. To address this challenge, a common approach is to treat noisy speech and clean speech as different domains and obtain invariant speaker embedding space through adversarial training \cite{8683828, 9054601}. Another approach is to learn feature representations that are independent of noise through disentanglement learning \cite{9413512, luu22_interspeech}. However, we find that noise disentangling can lead to the loss of some speaker-related information under clean conditions, resulting in poor performance of the SV system. In addition, few studies simultaneously consider extracting noise-independent speaker embeddings and establishing speaker invariant embedding spaces.

Inspired by this, we propose a noise disentanglement network architecture based on adversarial training to extract robust speaker embedding. Firstly, the disentanglement module includes a speaker encoder and a speaker-irrelevant encoder for decoupling speaker-relevant embedding and speaker-irrelevant embedding, respectively. The reconstruction component functions as a regularization constraint on the noise factor. And a feature-robust loss function guides the speaker encoder to learn noise-independent embeddings while preserving speaker information. In addition, adversarial training prevents speaker encoder from encoding various noisy information to promote model learning for more general representations. Experimental results confirm that our proposed method can achieve optimal performance under all conditions. 
\vspace{-0.2cm}
\section{Related Work}
\begin{figure*}[ht]
  \centering
  \includegraphics[width=\linewidth]{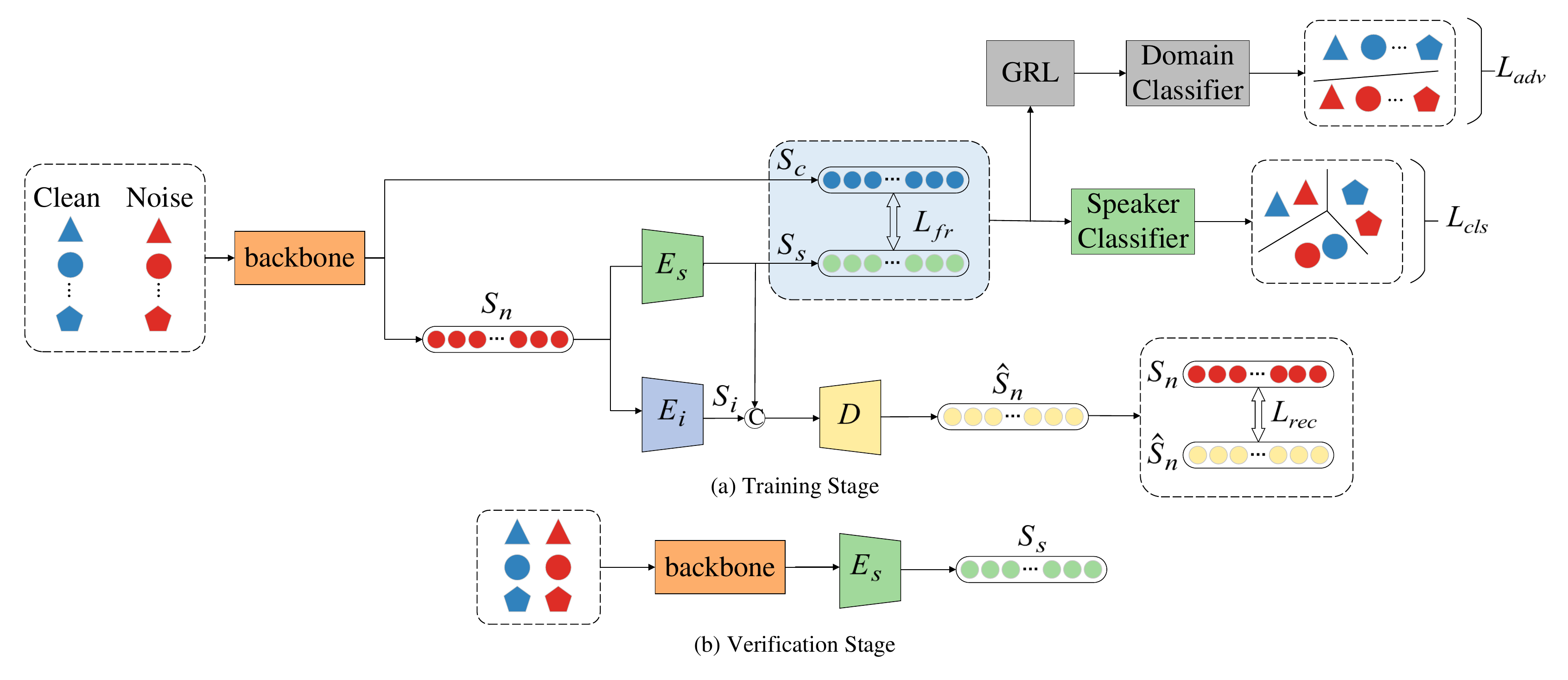}
  \caption{The architecture of noise-disentanglement adversarial training. Identical symbols correspond to the same speaker. The data depicted in blue and red denote the original and the augmented datasets, respectively.}
  \label{fig:ndal}
\end{figure*}
\vspace{-0.1cm}
\subsection{TDNN for deep speaker embedding}
\vspace{-0.1cm}
The most commonly used deep neural networks for extracting speaker embeddings are residual neural networks (ResNet) \cite{7780459}, time-delayed neural networks (TDNN) \cite{snyder17_interspeech}, or convolutional neural networks (CNN) \cite{chung18b_interspeech}. In this study, we used ECAPA-TDNN \cite{desplanques20_interspeech} to extract speaker embedding. In addition to applying statistics pooling to project variable-length utterances into fixed-length speaker embeddings, ECAPA-TDNN proposes further architecture enhancements to both the TDNN architecture and statistics pooling layer. Additional skip connections are introduced to propagate and aggregate channels throughout the system, and channel attention using global context is added to the frame layers and statistics pooling layer. Finally, the speaker embedding is extracted through a fully connected layer.

\vspace{-0.2cm}
\subsection{NDML-based method}
\vspace{-0.1cm}
Our method is related to the recently proposed Noise-Disentanglement Metric Learning (NDML) method \cite{10096848}, which is a SV system based on noise-disentanglement with metric learning to tackle the challenge of noise robustness under noisy environments. Inspired by NDML, we propose a novel noise-disentanglement network architecture based on multi-task adversarial training to achieve noise robustness. We will discuss their differences and emphasize the advantages of our approach in Section 3. 
\vspace{-0.2cm}
\section{Proposed Methods}
\vspace{-0.1cm}
The proposed noise-disentanglement based on adversarial training architecture consists of three modules: a backbone \(B\), a disentanglement module and an adversarial training module, as illustrated in Figure~\ref{fig:ndal}. The disentanglement module includes a speaker encoder \(E_s\), a speaker-irrelevant encoder \(E_i\) and a reconstruction module \(D\). And the adversarial training module, which includes a binary domain classifier with a gradient reversal layer, is used to discourage \(E_s\) from encoding acoustic condition information. The parameters of the backbone, speaker encoder, speaker-irrelevant encoder and decoder are accordingly denoted as \(\theta\), \(\phi _{s} \), \(\phi _{i} \) and \(\phi _{d} \), respectively. Finally, the reconstruction loss, feature-robust loss, classification loss and adversarial loss are used jointly to optimize the speaker encoder and backbone network.
\vspace{-0.1cm}
\subsection{Noise disentanglement}
\vspace{-0.1cm}
Noise can disrupt the voiceprint features of clean speech. To alleviate this, we propose the noise disentanglement for purifying clean speaker information from corrupted speech at the speaker embedding level. Different from the NDML \cite{10096848} that focuses on the disentanglement of feature level, which is susceptible to noise interference, better results can be achieved at the deep speaker embedding level decoupling.

Speaker encoder \(E_s\) and speaker-irrelevant encoder \(E_i\) are designed to capture speaker representation \(S_s\) and speaker-irrelevant representation \(S_i\) from the noisy speaker embedding \(S_n\), respectively. The reconstruction module serves as a constraint term to promote decoupling. Then, the concatenation of \(S_s\) and \(S_i\) serves as input to decoder \(D\) to reconstruct noisy speaker embedding \(S_n\). MSE loss is used to minimize the distance between \(S_n\) and \(\hat{S}_{n}\), as follows:
\begin{align}
  L_{rec} = \frac{1}{N} \sum_{i=1}^{N} ( S_{n}^{i} - \hat{S}_{n}^{i} )^{2} 
  \label{equation:eq2}
\end{align}
where \(N\) is the batch size.

A feature-robust loss between clean speaker embedding \(S_c\) and decoupled noisy embedding \(S_s\) is optimized to supervise the speaker encoder \(E_s\) to generate a noise-independent speaker embedding without losing speaker information, as follows: 
\begin{align}
  L_{fr} = \frac{1}{N} \sum_{i=1}^{N} ( S_{c}^{i} - {S}_{s}^{i} )^{2} 
  \label{equation:eq1}
\end{align}

Then, \(S_c\) and \(S_s\) are fed into the speaker classifier simultaneously, to calculate the classification loss using AAM-Softmax:
\begin{align}
  L_{cls} = - \frac{1}{2N} \sum_{i=1}^{2N} log\frac{e^{s\cdot cos(\theta _{y_{i}} + m)}} {e^{s\cdot cos(\theta _{y_{i}} + m)} + \sum_{j=1,j\ne y_{i} }^{C} e^{s\cdot cos(\theta _{j} )}}  
\end{align}
where \(y_{i}\) represents the speaker label of the i-th utterance, \(s\) and \(m\) are two hyperparameters for AAM-Softmax.

\begin{table*}[ht]
\caption{EER (\%) of various systems under the clean and seen noisy environment. Best in bold}
\label{tab:clean&seen}
\centering
\setlength{\tabcolsep}{7pt}
\begin{tabular}{ccccccccc}
\hline
\textbf{Noise Types}    & \textbf{SNR} & \textbf{Clean} & \textbf{Joint} & \textbf{Li et al. \cite{li21b_interspeech}} & \textbf{NDML \cite{10096848}} & \textbf{w/o AL} & \textbf{w/o Dis} & \textbf{NDAL} \\ \hline
Clean                   & -            & 3.99           & 3.64           & 4.57               & 2.90          & 2.74            & 2.82             & \textbf{2.63} \\ \hline
\multirow{5}{*}{Music}  & 0            & 20.52          & 9.89           & 11.82              & 10.84         & 7.13            & 7.07             & \textbf{6.43} \\ 
                        & 5            & 11.58          & 6.60           & 7.71               & 6.52          & 4.85            & 4.77             & \textbf{4.44} \\  
                        & 10           & 7.11           & 5.12           & 5.70               & 4.66          & 3.77            & 3.70             & \textbf{3.59} \\ 
                        & 15           & 5.16           & 4.13           & 5.00               & 3.67          & 3.33            & 3.21             & \textbf{3.08} \\ 
                        & 20           & 4.50           & 3.89           & 4.76               & 3.21          & 3.06            & 2.90             & \textbf{2.87} \\ \hline
\multirow{5}{*}{Noise}  & 0            & 17.71          & 8.84           & 9.49               & 10.24         & 6.24            & 6.48             & \textbf{5.87} \\
                        & 5            & 11.94          & 6.67           & 7.31               & 6.96          & 4.66            & 4.72             & \textbf{4.19} \\ 
                        & 10           & 8.37           & 5.23           & 6.12               & 5.02          & 3.67            & 3.77             & \textbf{3.53} \\
                        & 15           & 6.30           & 4.55           & 5.48               & 3.91          & 3.28            & 3.39             & \textbf{3.23} \\ 
                        & 20           & 4.98           & 4.03           & 5.04               & 3.40          & 3.10            & 3.12             & \textbf{3.09}  \\ \hline
\multirow{5}{*}{Speech} & 0            & 22.73          & 11.41          & 36.79              & 10.96         & 6.57            & 6.80             & \textbf{6.14} \\ 
                        & 5            & 12.72          & 6.58           & 19.27              & 6.13          & 4.44            & 4.43             & \textbf{4.00} \\ 
                        & 10           & 7.45           & 4.81           & 9.96               & 4.28          & 3.65            & 3.66             & \textbf{3.23} \\ 
                        & 15           & 5.45           & 4.16           & 6.74               & 3.52          & 3.05            & 3.13             & \textbf{2.97} \\ 
                        & 20           & 4.53           & 3.89           & 5.44               & 3.21          & 2.96            & 2.96             & \textbf{2.80} \\ \hline
Average                 & -            & 9.69           & 5.84           & 9.45               & 5.59          & 4.16            & 4.18             & \textbf{3.88} \\ \hline
\end{tabular}
\end{table*}

\subsection{Adversarial training}

However, noise-disentanglement does not fully separate speaker information from speaker-irrelevant information. To increase the degree of disentanglement and establish a speaker-invariant space, we propose adversarial training to discourage \(E_s\) from encoding acoustic condition information.

In order to utilize adversarial training in this case, we use augmentation labels (raw/augmented) instead of acoustic condition labels. Therefore, the domain classifier is designed as a binary classifier to maximize the correct prediction of augmentation labels for speaker embedding \(S_c\) and \(S_s\). And during backpropagation, the gradient reversal layer is used to force the backbone \(B\) and speaker encoder \(E_s\) to generate speaker embeddings independent of noise, making it impossible for the domain classifier to distinguish, thereby achieving a minimax game. The adversarial cost function \(L_{adv}\) is defined as the cross-entropy,
\begin{align}
  L_{adv} = - \frac{1}{2N} \sum_{i=1}^{2N} a_{i} \cdot log(Softmax(F(S_{a}^{i})))   
\end{align}
where \(a_{i}\) is the augmentation label of the i-th utternace, \(F\) is the domain classifier, and \(S_{a}\) is the set of \(S_c\) and \(S_s\). 

Through adversarial training, the backbone \(B\) and speaker encoder \(E_s\) can be maximally motivated to learn noise-independent speaker embeddings and achieve speaker invariant embedding space. The whole cost function \(L\) is formulated below:
\begin{align}
  L = L_{rec} + L_{fr} + L_{cls} - \lambda L_{adv}
\end{align}
where \(\lambda \) is a positive gradient reversal coefficient that controls the trade-off between multiple objectives during training process.

For each step,  \(\phi _{s}^{t} \) is updated to the value of \(\phi _{s}^{t+1} \) using reconstruction loss \(L_{rec}\), feature-robust loss \(L_{fr}\), classification loss \(L_{cls}\) and adversarial loss \(L_{adv}\), as follows:
\begin{align}
  \phi _{s}^{t+1} = \phi _{s}^{t} - \alpha \bigtriangledown _{ \phi _{s}^{t} } ( L_{rec} + L_{fr} + L_{cls} - \lambda L_{adv})
\end{align}
where \(\alpha\) is the learning rate.

\section{Experiments}

\subsection{Datasets}

Following the common experiment settings \cite{li21b_interspeech, 10096848}, experiments are conducted on the VoxCeleb1 \cite{nagrani17_interspeech} dataset. The development set contains 148642 utterances from 1211 speakers. And the test set contains 4874 utterances from 40 speakers, which constructs 37720 test trials. Since the dataset is collected in the wild, the speech segments are corrupted with real-world noise. But we assume the raw data to be a clean dataset and generate noisy data based on this raw data. The MUSAN \cite{snyder2015musan} dataset is used as the source of noise, which contains 60 hours of speech, 42 hours of music and 6 hours assorted noise. The MUSAN dataset is divided into two non-overlapping subsets for generating noisy training and testing utterances respectively. At the training stage, for each clean utterance, one noisy utterance is generated at the random SNR level from 0dB to 20dB with a random noise type. At the testing stage, we evaluate the performance of the SV systems under seen and unseen noisy environments. For the seen noisy environments, the noise data is sampled from the remaining half of the MUSAN dataset. For the unseen noisy environments, we use NoiseX-92 \cite{varga1993noisex} dataset and Nonspeech dataset as another noise source to generate noisy testing utterances. The NoiseX-92 dataset includes 15 kinds of noise, such as White Noise and Pink Noise. The nonspeech dataset consists of 100 types of noise, which is collected in various life scenarios.

\begin{figure*}[ht]
  \centering
  \includegraphics[width=\linewidth]{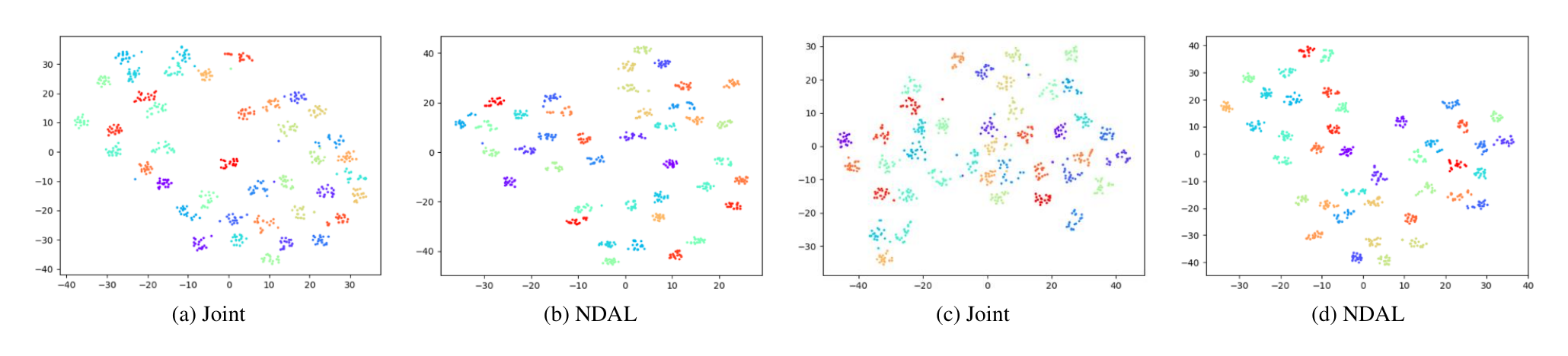}
  \caption{The t-SNE visualization of speaker embeddings: (a, b) visible noise vs. (c, d) invisible noise conditions.}
  \label{fig:tSNE}
\end{figure*}

\subsection{Implementation details}

The input features are 80-dimensional log mel spectrogram features from a 25 ms window with a 10 ms frame shift, which is normalized through cepstral mean subtraction and no voice activity detection is applied. During the training stage, 3s segments are randomly selected from each original utterance. Additionally, SpecAugment \cite{park19e_interspeech} is applied on the log mel spectrogram of the samples, where 0 to 10 channels in the frequency domain and 0 to 5 frames in the time domain are randomly masked. One clean and one noisy utterance per 150 randomly selected speakers, totaling 300 utterances, are grouped as one batch and fed into the systems. All systems are trained using Additive Angular Margin Softmax (AAM-softmax) with a margin of 0.2 and a scaling factor of 30, except that the loss function for the domain classifier is defined as cross-entropy. For optimization, the Adam optimizer with an initial learning rate of 0.001, a learning rate decay of 0.97 and the weight decay of 2e-5 is used to train the whole network.

ECAPA-TDNN network is used as the speaker embedding extractor for its simplicity, with 1024 channels in the convolutional frame layers. After training, the 192-dimensional speaker embeddings are extracted through the backbone and speaker encoder. The whole utterance is used to extract speaker embeddings during the test stage. The cosine similarity is used for scoring. And the equal error rate (EER) is used as the performance metric.

The speaker encoder and speaker-irrelevant are 2-layer AutoEncoders with hidden size of 1024. The decoder is almost the same as the encoders.

\begin{table}[ht]
\caption{EER (\%) of various systems under unseen noisy environment. Best in bold}
\label{tab:unseen}
\centering
\setlength{\tabcolsep}{4.5pt}
\begin{tabular}{cccccc}
\hline
\textbf{Unseen Noise}      & \textbf{SNR} & \textbf{Joint} & \textbf{w/o AL} & \textbf{w/o Dis} & \textbf{NDAL} \\ \hline
\multirow{5}{*}{NoiseX-92} & 0            & 13.50          & 9.36            & 9.55            & \textbf{9.07} \\ 
                           & 5            & 8.50           & 6.34            & 6.25             & \textbf{5.94} \\
                           & 10           & 6.33           & 4.84            & 4.76             & \textbf{4.52} \\
                           & 15           & 5.00           & 3.78            & 3.78             & \textbf{3.59} \\ 
                           & 20           & 4.23           & 3.24            & 3.30             & \textbf{3.17} \\ \hline
\multirow{5}{*}{Nonspeech} & 0            & 12.54          & 8.08            & 8.75             & \textbf{7.57} \\  
                           & 5            & 8.19           & 5.71            & 5.88             & \textbf{5.49} \\
                           & 10           & 6.03           & 4.29            & 4.31             & \textbf{4.03} \\
                           & 15           & 4.90           & 3.70            & 3.60             & \textbf{3.36} \\ 
                           & 20           & 4.32           & 3.08            & 3.21             & \textbf{2.99} \\ \hline
Average                    & -            & 7.35           & 5.25            & 5.34             & \textbf{4.97} \\ \hline
\end{tabular}
\end{table}
\vspace{-0.6cm}
\subsection{Results}

Table~\ref{tab:clean&seen} and~\ref{tab:unseen} show the performance under the seen and unseen noisy conditions, respectively. To observe the embedding distribution, we selected 40 speakers from the VoxCeleb1 test set, and randomly sampled 20 utterances from each speaker to generate speaker embeddings. The t-SNE visualization of speaker embeddings in visible and invisible noise conditions are plotted in Figure~\ref{fig:tSNE}. Clean means the baseline is trained on the original dataset. Joint means the baseline is trained on the original dataset and noisy dataset. As anticipated, the performance of the baseline, trained on the original dataset, markedly degrades in noisy environments. Data augmentation enhances the robustness of the model to noise. While Joint training surpasses clean training in effectiveness, the extent of this improvement is constrained. The model trained with noise-disentanglement metric learning (NDML) \cite{10096848} is used to compare.

Table~\ref{tab:clean&seen} illustrates that our method can generally achieve the best results under the clean and seen noisy conditions. In the average of overall conditions, the proposed NDAL achieves 33.56\% relative reduction in the terms of EER compared to the baseline joint training model. For clean scenarios, NDAL outperforms baseline 27.75\% in the EER. And our method has achieved better performance compared to NDML \cite{10096848}. Experimental results reveal that our method yields greater improvement in noisy environments, which is attributed to the robustness of our method to noise. In addition, optimizing feature-robust loss can effectively ensure that speaker related information is not lost under clean conditions while generating a noise independent speaker embedding, which is an essential part.

Table~\ref{tab:unseen} shows that our proposed method outperforms the baseline under the unseen noisy environments. Although Nonspeech dataset contains a wider variety of noise types compared to NoiseX-92 dataset, the performance of our model in these two unseen environments is essentially similar. This further demonstrates that our method can be robust to unseen noise. Due to the lack of prior knowledge of noise distribution, noise problems become more difficult for invisible environments \cite{7041826}. However, our model performs well in unseen environments, exhibiting strong generalization ability. On average, compared to the baseline, NDAL achieves 32.38\% relative reduction in EER. This performance improvement is attributed to our enhanced-disentanglement module and adversarial training approach, which enables the model to learn a speaker-invariant embedding space that is noise-independent. As shown in Figure~\ref{fig:tSNE}, in visible noise environments, our method can achieve better speaker embedding distribution compared to the baseline, which is more significant in invisible noise environments.
\vspace{-0.2cm}
\subsection{Ablation studies}
\vspace{-0.1cm}
The ablation study is conducted to evaluate the effect of the individual components in Section 3. NDAL (w/o AL) means that we only keep the disentanglement module. NDAL (w/o Dis) signifies that we train the baseline with adversarial training. NDAL (w/o AL) achieves 28.77\% and 28.57\% relative reduction in EER compared to baseline on average under both seen and unseen scenarios, respectively. And NDAL (w/o Dis) obtains 28.43\% and 27.35\% relative reduction in EER compared to baseline on average under both seen and unseen scenarios, respectively. It can be observed that disentanglement module and adversarial training play a crucial role in improving the system performance. Furthermore, the synergistic combination of these two approaches yields the most optimal performance.

\section{Conclusion}

In this work, we proposed a novel speaker verification system based on noise-disentanglement adversarial training to address the challenge of noise robustness under noisy environments. Specifically, the disentanglement module is used to capture noise-robust speaker embeddings. Adversarial training is used to discourage speaker encoder from encoding acoustic information, generating a speaker-invariant embedding space. Experimental results indicate that our method can enhance the robustness of SV system under both seen and unseen noisy conditions.  

\bibliographystyle{IEEEtran}
\bibliography{mybib}

\end{document}